\documentclass[reprint,amsmath,amssymb,prl,aps,superscriptaddress]{revtex4-1}
\UseRawInputEncoding
\usepackage{txfonts}
\usepackage{pifont}
\usepackage{amssymb}
\usepackage{dcolumn}
\usepackage{amsmath}
\usepackage[dvips]{epsfig}
\usepackage{array}
\usepackage{color}
\usepackage{ulem}

\makeatletter
\def\btt#1{\texttt{\@backslashchar#1}}%
\DeclareRobustCommand\bblash{\btt{\@backslashchar}}%
\makeatother

\begin{document}
\bibliographystyle{apsrev4-1}
\title{Superconductivity, nematicity, and charge density wave in high-T$_c$ cuprates: A common thread}

\author{Zhong-Bing Huang}
\email{huangzb@hubu.edu.cn}
\affiliation{Faculty of Physics and Electronic Technology, Hubei University, Wuhan 430062, China}
\affiliation{Beijing Computational Science Research Center, Beijing 100193, China}

\author{Shi-Chao Fang}
\affiliation{Faculty of Physics and Electronic Technology, Hubei University, Wuhan 430062, China}

\author{Hai-Qing Lin}
\affiliation{Beijing Computational Science Research Center, Beijing 100193, China}

\date{\today}

\begin{abstract}
To address the issues of superconducting and charge properties in high-T$_c$ cuprates, we perform a quantum
Monte Carlo study of an extended three-band Emery model, which explicitly includes attractive interaction
$V_{OO}$ between oxygen orbitals. In the physically relevant parameter range, we find that $V_{OO}$
acts to strongly enhance the long-range part of d-wave pairing correlation, with a clear tendency to form
long-range superconducting order in the thermodynamic limit. Simultaneously, increasing $|V_{OO}|$ renders
a rapid increase of the nematic charge structure factor at most of wavevectors, especially near
$\textbf{q}=(0,0)$, indicating a dramatic enhancement of nematicity and charge density waves. Our findings
suggest that the attraction between oxygen orbitals in high-T$_c$ cuprates is a common thread linking
their superconducting and charge properties.

\end{abstract}

\maketitle

\textit{Introduction--}
Superconductivity (SC) and mysterious pseudogap in high-T$_c$ cuprates have been the challenging issues
in the condensed matter physics
community~\cite{Keimer2015,Rice2012,Vojta2009,Lee2006,Damascelli2003,Varma2019,Casey2011,Chu2015}.
The pseudogap phase setting in at
the temperature T* harbours a rich variety of symmetry-broken orders~\cite{Keimer2015}.
Nematicity, a C$_4$ rotational-symmetry-broken order, is manifested by different in-plane anisotropic
properties~\cite{Ando2002,Hinkov2008,Sato2017,Li2021,Daou2010,Lawler2010} below the onset temperature
T$_{nem}$. The coincidence of T$_{nem}$ and T* observed in some experiments including transport
measurement~\cite{Daou2010} and scanning tunneling microscope~\cite{Lawler2010} strongly
suggests that nematicity has an intimate relation to the formation of pseudogap. Another emerging order, the
so-called charge density wave (CDW), prevalent in all chemically distinct families of cuprates, simultaneously
breaks the rotational and translational
symmetries~\cite{Frano2020,Comin2016,Chen2016,Gerber2015,Ghiringhelli2012}.
This order sets in between 100~K and 200~K and has a d-wave
form factor between O$_x$ and O$_y$ orbitals on the CuO$_2$ plane~\cite{Hamidian2016,Comin2015}.

Recently, Raman scattering spectroscopies of several copper oxides showed that with decreasing the hole doping
density, the pseudogap energy scale increases in a similar manner to the doping dependence of the CDW and
antinodal superconducting energy scales. In addition, the latter two have almost equal values over a substantial
doping region. These observations revealed that the three energy scales may be governed by a common thread,
and the CDW and d-wave SC are intimately related~\cite{Loret2019,Loret2020}.

Tremendous theoretical attempts have been made to understand the d-wave SC, nematicity and CDW in high-T$_c$ cuprates
based on the single- and multi-band models, but so far the answers remain highly controversial. First, both positive~\cite{Jiang2019,Jiang2018,Sakai2016,Corboz2014,Gull2013,Raghu2010,Andersen2010,Capone2006,
Maier2005,Senechal2005,Sorella2002,Kuroki1996}
and negative~\cite{Qin2020,Alexandrov2011,Aimi2007,Pryadko2004,Shih1998,Zhang1997A,Assaad1994} evidence was found
for the existence of d-wave SC, depending on the
computational methods, all of which have their pros and cons. Second, although nematicity could be induced
either by a d-wave Pomeranchuk instability~\cite{Slizovskiy2018,Halboth2000} or via quantum melting of charge
stripes~\cite{Kivelson2003,Kivelson1998}, its stability is
limited at temperatures about one order of magnitude lower than T$_{nem}$ and/or in a narrow filling region near
the van Hove singularity~\cite{Slizovskiy2018,Halboth2000,Zheng2014}.
Moreover, it remains controversial on the main contribution to the formation of nematicity and CDW, with possible
candidates including the off-site Coulomb repulsion between oxygen sites~\cite{Bulut2013,Fischer2011},
the on-site Coulomb repulsion at the copper site~\cite{Zegrodnik2020}, and spin fluctuations~\cite{Yamakawa2015,Thomson2015}.
Given the robustness of d-wave SC, nematicity and CDW in high-T$_c$ cuprates, divergent theoretical conclusions
suggest that a key ingredient may be missing in the microscopic models.

In this Letter, we study the superconducting and charge properties in the realistic three-band Emery
model~\cite{Emery1987} extended by an effective attraction $V_{OO}$ between oxygen orbitals. The introduction
of $V_{OO}$ is justified by a recent angle-resolved photoemission experiment~\cite{Chen2021} on the
one-dimensional cuprate Ba$_{2-x}$Sr$_x$CuO$_{3+\delta}$ across a wide range of hole doping, showing an
anomalously strong ¡°holon folding¡± branch near k$_F$. A comparison with the theory reveals there exists
a strong nearest-neighbor (NN) attraction of the order of $eV$ in the studied compound. Due to structural
and chemical similarities among cuprates, such a NN attraction may also be applicable to the 2D high-T$_c$
cuprates. Physically, $V_{OO}$ can arise from coupling to certain bosonic excitations, such as quantized waves
of electronic polarization~\cite{Mallett2013} and phonons~\cite{Wang2021,Huang2011,Huang2003}.

Our main results were obtained by the constrained-path Monte Carlo (CPMC) method~\cite{Zhang1995,Zhang1997B,Qin2016},
which can precisely capture the ground-state information of correlated electron systems. As expected, since
attractive $V_{OO}$ favors the hole pairing between oxygen orbitals, it induces a strong enhancement of the
long-range part of d-wave pairing correlation. Unexpectedly, $V_{OO}$ gives rise to an enhancement of the nematic
charge structure factor at most of wavevectors, and the enhancement effect is particularly strong near
$\textbf{q}=(0,0)$. This indicates that $V_{OO}$ benefits the formation of $\textbf{q}=(0,0)$ nematicity and
finite $\textbf{q}$ CDW. Our findings demonstrate that the attraction between oxygen orbitals is crucial for
a universal understanding of d-wave SC, nematicity and CDW in high-T$_c$ cuprates.

\textit{Model and computational method--}
To understand the physics of CuO$_2$ plane, we adopt the following Hamiltonian,
\begin{equation}
H=H_{Emery}+H_{attract},
\end{equation}
where $H_{Emery}$ and $H_{attract}$ stand for the standard three-band Emery model~\cite{Emery1987} and attractive
interaction between oxygen orbitals, respectively. They are expressed in the form:
\begin{eqnarray}
H_{Emery}&=&\sum_{\langle i,j\rangle\sigma}t_{pd}^{ij}
(d_{i\sigma}^{\dagger}p_{j\sigma}+h.c.)
+\epsilon\sum_{j\sigma}n_{j\sigma}^{p}\nonumber\\
&&+\sum_{\langle j,k\rangle\sigma}
t_{pp}^{jk}(p_{j\sigma}^{\dagger}
p_{k\sigma}+h.c.)+U_{d}\sum_{i}n_{i\uparrow}^{d}n_{i\downarrow}^{d},\nonumber\\
\end{eqnarray}
\begin{equation}
H_{attract}=V_{OO}\sum_{\langle\langle j,k\rangle\rangle}n_{j}^{p}n_{k}^{p},
\end{equation}
Here, $d_{i\sigma}^{\dagger}$ is the creation operator of a Cu $3d_{x^2-y^2}$ hole and $p_{j\sigma}^{\dagger}$
is the creation operator of an O $2p_x$ or $2p_y$ hole. $t_{pd}^{ij} = \pm t_{pd}$ and $t_{pp}^{jk} = \pm t_{pp}$
are the NN Cu-O and O-O hybridizations, respectively, with the Cu and O orbital phase factors included
in the sign. $U_d$ and $\epsilon$ denote the Coulomb repulsion at the Cu site and the charge-transfer energy.
$V_{OO}$ stands for attraction between O sites separated by a Cu site, and the corresponding summation restriction
is denoted by $\langle\langle j,k\rangle\rangle$.
In units of $t_{pd}$, we choose a physically relevant parameter set
$U_{d}=6.0$, $\epsilon=3.0$, and $t_{pp}=0.5$, while $V_{OO}$ is varied from $0.0$ to $-0.6$.

Our calculations were performed on the square lattices of $N=L \times L$ unit cells with periodic boundary
conditions imposed using the CPMC method~\cite{Zhang1995,Zhang1997B,Qin2016}. The basic strategy of CPMC is to
project out the ground-state wave function from an initial wave function by branching random walk in an
overcomplete space of constrained Slater determinants, which have positive overlaps with a known trial wave
function. In this work, we focus on the closed-shell case, for which the corresponding free-electron wave
function is nondegenerate and translationally invariant. In this case, the free-electron wave function
is a good choice as the trial wave function~\cite{Zhang1997B,Guerrero1998,Huang2001A}.

The d-wave pairing correlation is defined as:
\begin{equation}
\label{pairing}
C(\textbf{R}) = \langle\Delta_{d}^\dagger(\textbf{R}) \Delta_{d}(\textbf{0})\rangle,
\end{equation}
where
\begin{eqnarray*}
\Delta_{d}(\textbf{R}) &=& \sum\limits_{\boldsymbol\mu}f_{d}
({\boldsymbol\mu}) \{[d_{\textbf{R}\uparrow}d_{\textbf{R}+{\boldsymbol\mu}
\downarrow} -d_{\textbf{R}\downarrow}d_{\textbf{R}+{\boldsymbol\mu}\uparrow}]\\
&&+ [p^x_{\textbf{R}\uparrow}p^x_{\textbf{R}+{\boldsymbol\mu}\downarrow}
-p^x_{\textbf{R}\downarrow}p^x_{\textbf{R}+{\boldsymbol\mu}\uparrow} ]\\
&&+[p^y_{\textbf{R}\uparrow}p^y_{\textbf{R}+{\boldsymbol\mu}\downarrow}
-p^y_{\textbf{R}\downarrow}p^y_{\textbf{R}+{\boldsymbol\mu}\uparrow}] \}
\end{eqnarray*}
with ${\boldsymbol\mu}=\pm \hat{x},~\pm \hat{y}$. The d-wave form factor
$f_{d}({\boldsymbol\mu}) = 1 $ for ${\boldsymbol\mu} = \pm \hat{x}$
and $f_{d}({\boldsymbol\mu}) = -1 $ for ${\boldsymbol\mu} = \pm \hat{y} $.
The vertex contribution to the pairing correlation is given by,
\begin{equation}
V(\textbf{R}) = C(\textbf{R}) - {C'}(\textbf{R}),
\end{equation}
where ${C'}(\textbf{R})$ is the bubble contribution from the dressed (interacting) propagator~\cite{White1989}.
For convenience, we also define the partial averages of pairing correlation and its vertex contribution:
$\bar{C}(R>2.0)=(1/N')\sum_{R>2.0}C(\textbf{R})$ and $\bar{V}(R>2.0)=(1/N')\sum_{R>2.0}V(\textbf{R})$, where $N'$
is the number of hole pairs with $R>2.0$.

To understand the charge property, we compute the nematic charge correlation $N^{d}(\textbf{R})$ and corresponding
structure factor $N^{d}(\textbf{q})$, which are defined as,
\begin{equation}
\label{NdR}
N^{d}(\textbf{R}) = \langle (n^{px}_{\textbf{R}}-n^{py}_{\textbf{R}})(n^{px}_{\textbf{0}}-n^{py}_{\textbf{0}})\rangle,
\end{equation}
\begin{equation}
\label{Ndq}
N^{d}(\textbf{q}) = \sum_{\textbf{R}} e^{i\textbf{q}\cdot\textbf{R}}N^{d}(\textbf{R}).
\end{equation}
Here, the minus sign in Eq.~(\ref{NdR}) reflects the d-wave form factor or $\pi$-phase difference of charge
distribution between O$_x$ and O$_y$ orbitals.

\textit{Superconducting property--}
\begin{figure}
\centering
\includegraphics[scale=0.5]{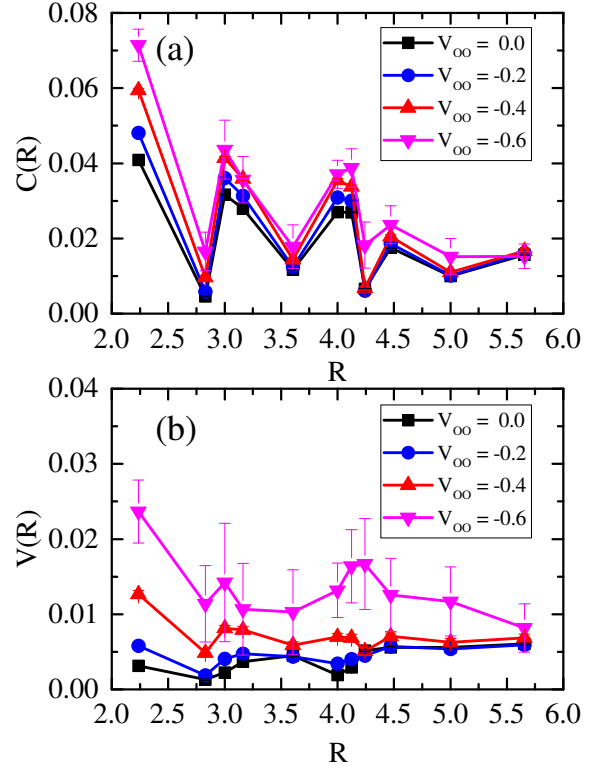}
\caption{(color online) (a) D-wave pairing correlation $C(R)$ and (b) vertex contribution $V(R)$
as a function of the distance $R$ on the $8 \times 8$ lattice at a hole doping density $\delta=0.156$.
The value of $V_{OO}$ is indicated by the shape of symbol.
}\label{dwave_L8}
\end{figure}
In Figs.~\ref{dwave_L8}(a) and \ref{dwave_L8}(b) we show $C(R)$ and $V(R)$ as a function of $R$ and $V_{OO}$ on
the $8 \times 8$ lattice at a hole doping density $\delta=0.156$. At $V_{OO}=0.0$, both $C(R)$ and $V(R)$ take
small positive values, which are consistent with previous results on the $6 \times 4$~\cite{Guerrero1998}
and $6 \times 6$~\cite{Huang2001A} lattices.
With decreasing $V_{OO}$ from $0.0$ to $-0.6$, one can clearly see that both pairing correlation and its vertex
contribution monotonically increase at all long-range distances for $R > 2.0$. Our simulations demonstrate that
the attractive interaction between oxygen orbitals is favorable for the d-wave SC. This is in sharp contrast to
the situation of one-band $t-U-V$ Hubbard model, where a NN attractive $V$ was expected to enhance
the d-wave SC~\cite{Micnas1990}, however, numerical studies indicated that the d-wave pairing correlation can
hardly be enhanced by $V$~\cite{Huang2001B,Nazarenko1996}.

\begin{figure}
\centering
\includegraphics[scale=0.5]{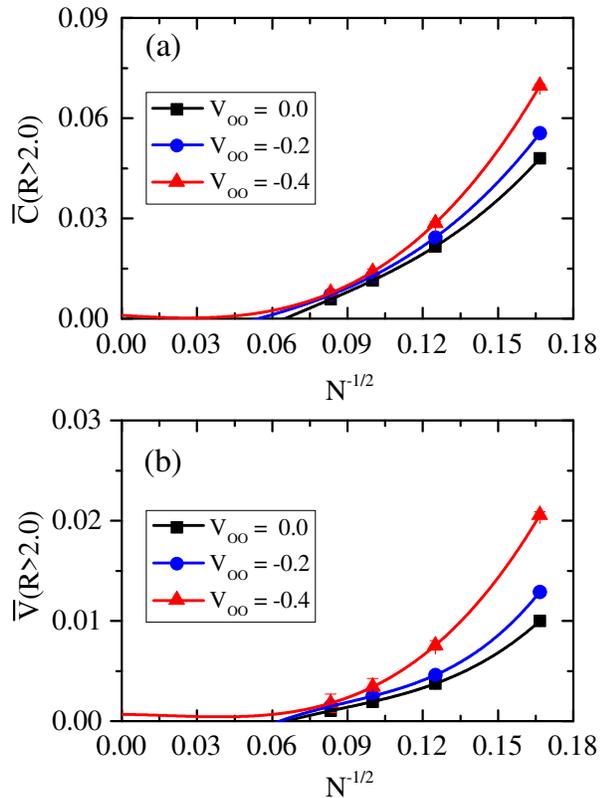}
\caption{(color online) (a) Partial average of pairing correlation $\bar{C}(R>2.0)$ and (b) partial average of
vertex contribution $\bar{V}(R>2.0)$ as a function of $1/\sqrt{N}$ for different values of $V_{OO}$.
The data for a fixed $V_{OO}$ are fitted by using third order polynomials in $1/\sqrt{N}$.
}\label{dwave_NL}
\end{figure}
To determine whether the d-wave SC exists in the extended three-band Emery model, we examine the evolution of
pairing correlation and its vertex contribution with the lattice size.
In Fig.~\ref{dwave_NL} we plot $\bar{C}(R>2.0)$ and $\bar{V}(R>2.0)$ as a function of $1/\sqrt{N}$ and $V_{OO}$.
The data were collected from the lattices with $N = 36,~64,~100$ and $144$, and the corresponding hole doping
densities are $0.167,~0.156,~0.140$ and $0.152$, respectively. For a fixed $V_{OO}$, the numerical results are
fitted by the third polynomial function of $1/\sqrt{N}$. When $V_{OO}=0.0$ and $-0.2$, the fitted curves touch the
horizontal axis at positive values, suggesting absence of long-range correlation between hole pairs in the
thermodynamic limit ($1/\sqrt{N}\rightarrow 0$). As $V_{OO}$ is decreased to $-0.4$, the positive intersections
with the vertical axis for both $\bar{C}(R>2.0)$ and $\bar{V}(R>2.0)$ clearly point towards the presence of
long-range superconducting order in the thermodynamic limit. We would like to point out that the accuracy of
extrapolation of our data to the thermodynamic limit is indeed affected by slightly different hole doping
densities on different lattices, as well as limited lattice sizes accessible to CPMC simulations, nevertheless
the monotonic increase of pairing correlation with decreasing $V_{OO}$ strongly supports the existence of d-wave
SC beyond certain $V_{OO}$.

\textit{Nematicity and CDW--}
\begin{figure}
\centering
\includegraphics[scale=0.5]{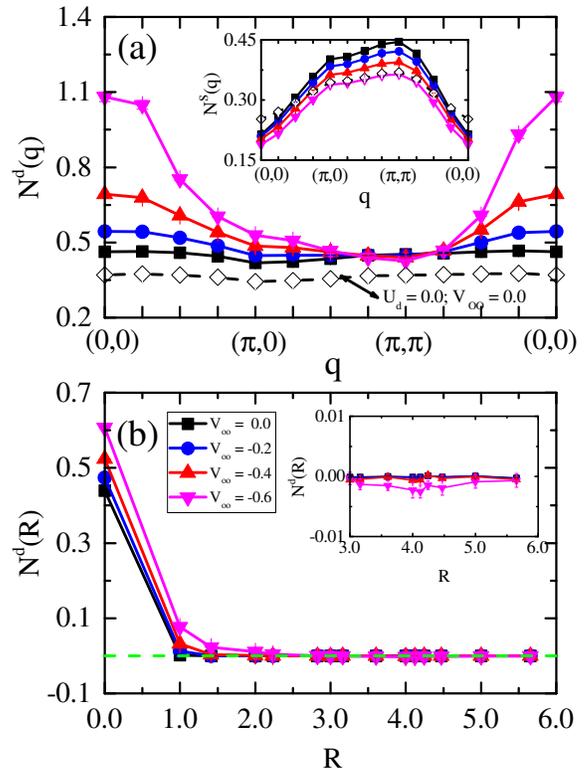}
\caption{(color online) (a) Nematic charge structure factor $N^{d}(\textbf{q})$ for different values of
$V_{OO}$ along the high symmetry lines in the first Brillouin zone. The open diamonds indicate the nematic
charge structure factor of free-electron system. Inset: The charge structure factor $N^{s}(\textbf{q})$
with the s-wave form factor as a function of $\textbf{q}$. (b) Nematic charge correlation $N^{d}(R)$ for
different values of $V_{OO}$ as a function of the distance $R$. Inset: The enlarged nematic charge
correlation for $R\ge 3.0$. The lattice size and hole doping density are the same as used in Fig.~\ref{dwave_L8}.
}\label{nematic_L8}
\end{figure}
Next, we turn to discuss the effect of $V_{OO}$ on the charge property. Fig.~\ref{nematic_L8}(a) displays
$N^{d}(\textbf{q})$ as a function of wavevector $\textbf{q}$ and $V_{OO}$ on the $8 \times 8$ lattice at
$\delta=0.156$. At $V_{OO}=0.0$, the shape of $N^{d}(\textbf{q})$ is very similar to the one of free-electron
system ($U_{d}=0.0$), without obvious peak appearing in the curve.
With decreasing $V_{OO}$ from $0.0$ to $-0.6$, $N^{d}(\textbf{q})$ exhibits a monotonic increase at all
$\textbf{q}'s$ except $(\pi,~\pi)$ and its nearby wavevectors, and this enhancement effect is particularly strong
near $\textbf{q}=(0,~0)$. The peak at $\textbf{q}=(0,~0)$ corresponds to a state in which the charge on the
$p^x$ orbital of each unit cell is larger than the one on the $p^y$ orbital, or vice versa. This sate is the
so-called nematic phase. Meanwhile, the enhancement at nonzero $\textbf{q}'s$ signals the development of CDW
with the d-wave form factor. It is worth noting that $N^{d}(\textbf{q})$ appearing at $(\pi/2,~0)$ is consistent
with the experimentally observed CDW peaked at ($[0.4\pi-0.6\pi],~0$)~\cite{Frano2020,Comin2016},
and stronger instability to charge ordering
at $\textbf{q}=(0,~0)$ than at $\textbf{q}=(\pi/2,~0)$ naturally accounts for why nematicity sets in preceding
the formation of CDW. As a comparison, we show in the inset of Fig.~\ref{nematic_L8}(a) the charge structure
factor with the s-wave form factor, $N^{s}(\textbf{q})$, obtained by replacing $-$ with $+$ in Eq.~(\ref{NdR}).
One can readily see that decreasing $V_{OO}$ leads to a suppression of $N^{s}(\textbf{q})$ at all $\textbf{q}'s$,
indicating that $V_{OO}$ is not beneficial for the formation of CDW with the s-wave form factor.

Fig.~\ref{nematic_L8}(b) shows the nematic charge correlation $N^{d}(R)$ as a function of $R$ and $V_{OO}$.
It is clear to see that only the short-range part of $N^{d}(R)$ with $R<3.0$ is enhanced by $V_{OO}$,
while the long-range part is close to zero and unaffected by $V_{OO}$ within simulation accuracy,
as shown in the inset figure. The rapid decrease of $N^{d}(R)$ to vanishingly small values indicates that
$N^{d}(\textbf{q})$ in Fig.~\ref{nematic_L8}(a) is contributed mainly from short-range nematic charge correlations
and there is no long-range charge order in the studied model.

\begin{figure}
\centering
\includegraphics[scale=0.5]{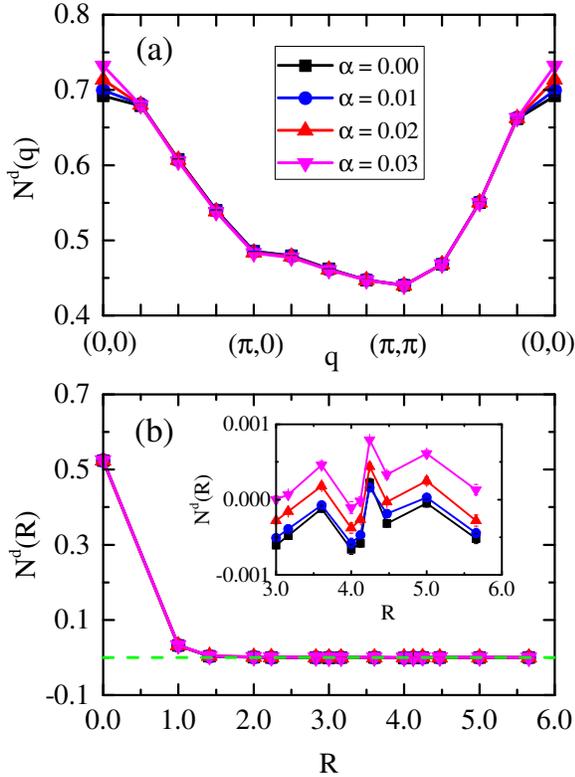}
\caption{(color online) (a) Nematic charge structure factor $N^{d}(\textbf{q})$ at $V_{OO}=-0.4$ for different
values of $\alpha$ along the high symmetry lines in the first Brillouin zone. (b) Nematic charge correlation
$N^{d}(R)$ at $V_{OO}=-0.4$ for different values of $\alpha$ as a function of the distance $R$.
Inset: The enlarged nematic charge correlation for $R\ge 3.0$. The lattice size and hole doping density are
the same as used in Fig.~\ref{dwave_L8}.
} \label{nematic_alpha}
\end{figure}
In high-T$_c$ cuprates, the C$_4$ rotational symmetry of CuO$_2$ plane can be broken by a transition from
tetragonal to orthorhombic structure. To address the effect of symmetry breaking, we mimic the orthorhombic
distortion by modifying the Cu-O hybridization as: $|t_{pd}^{ij}| = (1\pm\alpha)t_{pd}$, where $+(-)$ applies
if the Cu-O bond shrinks (stretches) by $\alpha$. Fig.~\ref{nematic_alpha} displays the $\textbf{q}$-dependent
$N^{d}(\textbf{q})$ and $R$-dependent $N^{d}(R)$ for different $\alpha$ on the $8 \times 8$ lattice
at $\delta=0.156$ and $V_{OO}=-0.4$. We observe that increasing $\alpha$ from $0.00$ to $0.03$ results in an
enhancement of $N^{d}(\textbf{q})$ at $\textbf{q}=(0,~0)$, but does not affect other ones at nonzero
$\textbf{q}'s$. The overlap of curves for different $\alpha$ in Fig.~\ref{nematic_alpha}(b) suggests that the
change of $N^{d}(R)$, if exists, is much smaller than the scale of the vertical axis. As seen from the inset
of Fig.~\ref{nematic_alpha}(b), the orthorhombic distortion induces a weak but significant enhancement of
long-range part of $N^{d}(R)$, which is crucial for the formation of long-range ordered nematic phase.
Based on the results presented in Fig.~\ref{nematic_alpha}, we can conclude that the orthorhombic distortion
is important for nematicity, but has little effect on the CDW. On the other hand, almost $\alpha$-independent
$C(R)$ (not shown here) indicates that the d-wave SC is insensitive to the structure symmetry.

\textit{Conclusion--}
We have studied the superconducting and charge properties of high-T$_c$ cuprates within the extended
three-band Emery model by using the CPMC method. The simulation results show that it is hard to establish the
d-wave SC and charge order in the standard three-band Emery model. Upon turning on the attraction
between oxygen orbitals, we observe a strong enhancement for both d-wave pairing correlation and nematic charge
structure factor at zero and finite wavevectors, which leads to simultaneous development of d-wave SC, nematicity,
and CDW. This finding provides a natural explanation for recent Raman scattering experiments,
where it was found that the energy scales of d-wave SC, CDW, and pseudogap seem to be governed by a common microscopic
interaction~\cite{Loret2019,Loret2020}. We expect that finite temperature studies of the extended three-band
Emery model will offer a deeper understanding of high-T$_c$ cuprates in the superconducting and pseudogap regions.

This work was supported by the National Natural Science Foundation of China (No.~11674087 and No.~11734002).
We acknowledge partial support from NSAF U1930402.


\end{document}